\documentclass[aps,prb,twocolumn,groupedaddress]{revtex4}
\usepackage{graphicx}
\usepackage{epsfig}
\begin{document}
\title{Magnetic-field effects on transport in carbon nanotube junctions}
\author{L. Rosales$^\ast$,  M. Pacheco$^\ast$, Z.
Barticevic$^\ast$, C. G. Rocha$^+$, and A.
Latg\'e$^\dag$\cite{email}} \affiliation{$^\ast$Departamento de
F\'\i sica, Universidad T\'ecnica F. Santa Maria, Casilla postal 110
V, Valpara\'iso, Chile} \affiliation{$^+$ Physics Institute, Trinity
College, Ireland} \affiliation{$^\dag$Instituto de F\'\i sica,
Universidade Federal Fluminense, 24210-340 Niter\'oi, RJ- Brazil}

\date{\today}

\begin{abstract}
Here we address a theoretical study on the behaviour of electronic
states of heterojunctions and quantum dots based on carbon nanotubes
under magnetic fields. Emphasis is put on the analysis of the local
density of states, the conductance, and on the characteristic curves
of current versus voltage. The heterostructures are  modeled by
joining  zigzag tubes  through single pentagon-heptagon pair
defects, and described within a simple tight binding calculation.
The conductance is calculated using the Landauer formula in the
Green functions formalism. The used theoretical approach
incorporates the atomic details of the topological defects by
performing an energy relaxation via Monte Carlo calculation. The
effect of a magnetic field on the conductance gap of the system is
investigated and compared to those of isolated constituent tubes. It
is found that the conductance gap of the studied CNHs  exhibits
oscillations as a function of the magnetic flux. However, unlike the
pristine tubes case, they are not Aharonov-Bohm periodic
oscillations.

\end{abstract}
\maketitle

\section{Introduction}

Following the richness possibilities explored first by the
semiconducting physicists, heterostructures made of carbon nanotubes
(CNs) have recently also been studied. In particular, the
combination of two or more kind of pristine tubes offers a variety
of physical situation mainly due to the intrinsical feature of the
carbon tubes, which exhibit electronic properties dictated by
geometrical aspects\cite{Dresselhaus}. This fact, together with
quite important mechanical characteristics, make clear that CNs may
be used in different devices in science and
nanotechnology\cite{Tans,Choi,Wie}. The presence of topological
defects can change the chirality of the CNs. It was shown that local
curvature and tube diameter do not suffer a drastic change when the
defect is a pentagon-heptagon pair\cite{Dunlap,Charlier}. Actually,
with this kind of defect it is possible to join two different CNs
forming a heterostructure similar to the semiconducting ones largely
studied\cite{TAndo}. Metal-metal and
metal-semiconducting\cite{Ferreira,Chico1,Ouyang,Hu,Fa} systems may
be naturally formed besides the standard semiconducting composites.
Of course, the electronic nature of each one of the CN components
and their symmetries will define the electronic and transport
properties exhibited by the resulting
heterostrutures\cite{Chico2,Dresse,Tamura,Yao,Yang,Farajian}. For
zigzag CNs, a change in one unity in the chiral number n [(n-1, 0)
or (n+1, 0)] leads to an electronic changing from semiconductor to
metallic electronic behavior and vice-versa, involving a small
change in the diameter of the tubes. Similar effect may be achieved
by applying a magnetic field to the carbon nanotube heterostructures
(CNH's), showing novel electronics and transport
behaviors\cite{Nakanishi,Lee,Dresse3,Fujiwara,Rocha}. By scanning
tunneling spectroscopy measurements at selected locations of CNs it
is possible to obtain a map of the electronic density of states.
This technique allowed the characterization of interface states
induced by the presence of defects at the junctions of two
semiconducting nanotubes\cite{Kim}, and also the determination of
spatial oscillation in the electronic density of states with the
period of atomic lattice.

Here we explore the gap energy modulation of nanotube
heterostructures (single junctions and quantum dots) under the
action of an external magnetic field. Emphasis is put on the
transport response dependence. Single junctions have been proved to
be stable and in particular the stability of a (7,0)/(6,0) CNH is
verified, in the absence of magnetic fields, by calculating the
total energy of the system via a numerical Monte Carlo analysis.
Fig.1 shows the atomic configuration of this particular
heterostructure, adopting a Tersoff empirical interatomic
potential\cite{Tersoff}. The topological defect (a pair of
pentagon-heptagon) is marked with bold lines (blue on-line). Despite
of the simplicity of the potential-model, it has been properly used
for carbon-based materials in the determination of total and defect
energy deviations, and elastic properties.  The resulted equilibrium
atomic position help us in determining the Peierl's phases when
considering the magnetic field since they depend on each one of the
atomic positions.
\begin{figure}[h!]
\begin{center}
\includegraphics[width=6.0cm]{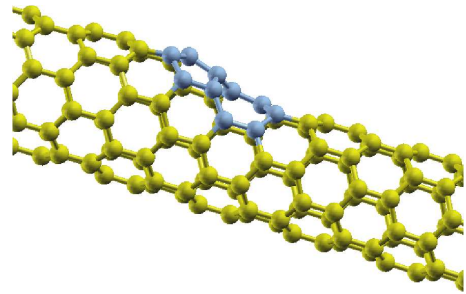}
\end{center}
\caption{''Color online''. Atomic configuration of a zigzag CNH,
(7,0)/(6,0) obtained by a Monte Carlo simulation using a
semiclassical interatomic potential.}
\end{figure}

To investigate how the magnetic field changes the electronic
properties of the heterostructures, we  calculate local density of
states (LDOS) and electrical conductance. We consider a single
$\pi$-band tight binding Hamiltonian and follow real-space
renormalization techniques. The conductance is calculated using the
Landauer formula in the Green function formalism\cite{Nardelli}. We
restrict our discussion to zigzag CNs junctions formed by the
presence of the pair defects, as shown in figure 1.  The external
magnetic field is considered uniform and parallel to the system
axis. The occurrence of the Aharonov-Bohm effect is investigated in
the composed nanostructures.

\section{Theoretical method }

We restrict our present study to (n,0)/(n$\pm$1,0) carbon nanotube
junctions (CNHs), and adopt a single $\pi$-band tight binding
Hamiltonian, taking into account a fixed value for the hopping
parameter ($\gamma_o \approx $ 2.75 eV), independent of the
orientation, location and length of the bond. The systems are
described in a real space picture which allow us to incorporate the
potential fluctuations at the microscopic scale. LDOS and
conductance of the structures are calculated within the Green
function formalism, employing decimation
procedures\cite{Ferreira,Rocha} (or, equivalently, ad-layers
schemes).
\begin{figure}[]
\begin{center}
\includegraphics[width=8.0cm,height=6.2cm]{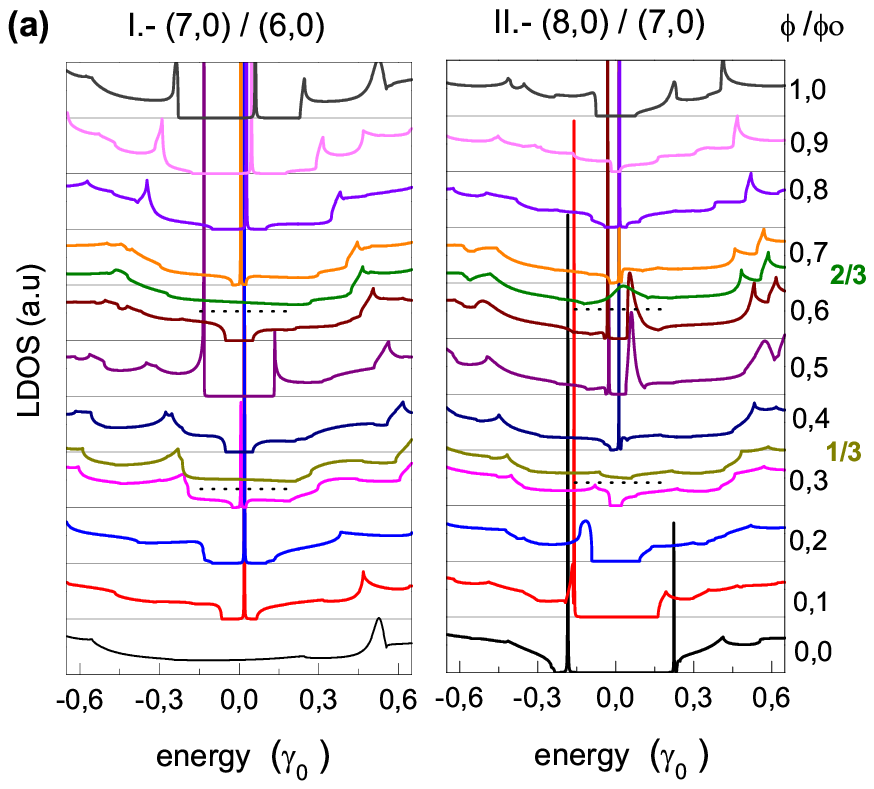}
\includegraphics[width=8.0cm,height=6.2cm]{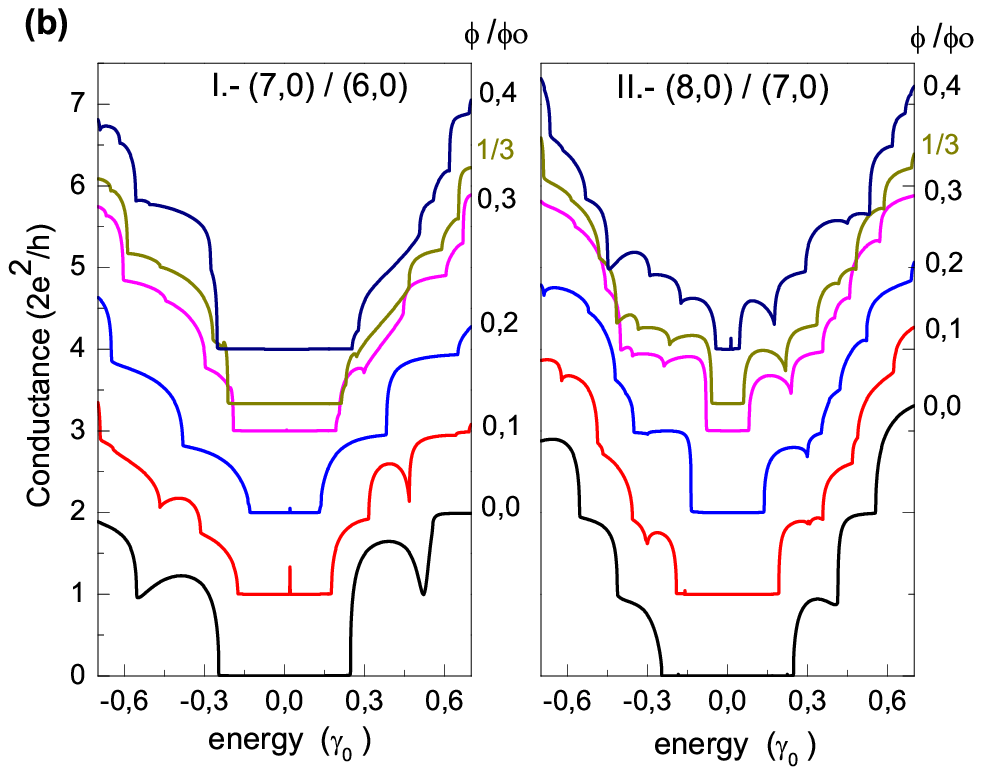}
\end{center}
\caption{''Color online''. (a) LDOS at the defective ring, (b)
conductance as function of energy for two types of single junctions:
I- Semiconductor/Metal CNH (7,0)/(6,0) and II- Semiconducting CNH
(8,0)/(7,0), for different values of magnetic flux. The curves are
shifted upwards for a better visualization.}
\end{figure}
The LDOS at site i is obtained directly from the renormalized
locator G$_{i,i}$ [$\rho_{ii}(\omega)=-1/\pi
Imag(Tr(G_{ii}(\omega)))$]. The surface Green´s functions matching
formalism is used to obtain the conductance combined with an
iterative calculation of transfer matrices\cite{Nardelli}. Within
this picture the full system is partitioned into three parts: the
central one and two leads composed of two carbon nanotubes. The
conductance is related to the scattering properties of the region
via the Landauer formula. In the linear response approach it can be
written in term of the Green's function of the system
by\cite{Garcia}

\begin{equation}
\Gamma (E_F)= \frac{2e^2}{h} T(E_F)\,\
\end{equation}
where $T(E_F)$  is the transmission function of an electron crossing
through a central conductor, given by  $T(E_F)= Tr(\Omega_{L} G_c^R
\Omega_R G_c^A)$, with Tr being the matricial trace function,
$G_c^{R,A}$ the retarded and advanced Green functions corresponding
to the central part of the system, and with
\begin{equation}
\Omega_{L,R}=i[\Sigma_{L,R}^R-\Sigma_{L,R}^A] \,\,\; \,\,\
\Sigma_{L,R}=V_{c,L/R}g_{L,R}V_{L/R,c}\,\,,
\end{equation}
describing the coupling between the central part and the right and
left leads, given by the corresponding self-energies. Here the
contacts are given by surface Green functions corresponding to the
(n,0) and (n$\pm$ 1,0) tubes whereas the conductor is the defective
ring. All the Green functions are obtained numerically and the
effects of the magnetic field are described within the Peierl's
phase approximation. In this scheme the hopping energies  are
modified by a phase which depends on the potential vector associated
with the field and on the relative atomic distances. One should
stress that the atomic positions of the structure, including the
defect region, were carefully studied using Tersoff~\cite{Tersoff}
relaxation process. In what follows, the energies are written in
terms of the hopping parameter $\gamma_o$, the magnetic fluxes in
units of the quantum flux $\phi_o$, and taking into account the flux
through the biggest tube of the heterostructures. The Fermi level
was taken as the zero of the energies.

\section{Results}
Two kinds of structures are considered in this work: a single
junction and a semiconductor nanotube quantum dot (CNQD). Both are
formed by the presence of the pentagon-heptagon defects. Results for
their electronic and transport properties are shown in the following
subsections where we discuss the effects on a magnetic field
threading the structure.

\subsection{Single junctions}

CNH's of type (n,0)/(n$\pm$1,0) allow us to consider two types of
junctions: semiconductor-metal (S/M) and semiconductor-semiconductor
(S/S) configurations. Results for LDOS and conductance for a S/M
[(7,0)/(6,0)] and a S/S [(8,0)/(7,0)] junctions are displayed in
Fig.2, for different magnetic field intensities. The plotted LDOS
are mean values calculated at the defective ring at which the
constituent tubes change their diameter. At zero field, the LDOS of
the S/M junction [(7,0)/(6,0)] exhibits a plateau close to the Fermi
level, whereas the S/S [(8,0)/(7,0)] junction essentially retains
the gap of the pristine (7,0). The topological defect produces the
apparition of interface states in the LDOS of the S/S junction,
localized close to the gap edge\cite{Charlier,Ferreira,Rocha,Liu}
[see figure 2(a)]. Both studied junctions present a wide gap in the
conductance produced basically by the lost of the rotational
symmetry in the defective ring\cite{Chico2,Rocha}.
\begin{figure}[h!]
\begin{center}------
\includegraphics[width=6.5cm]{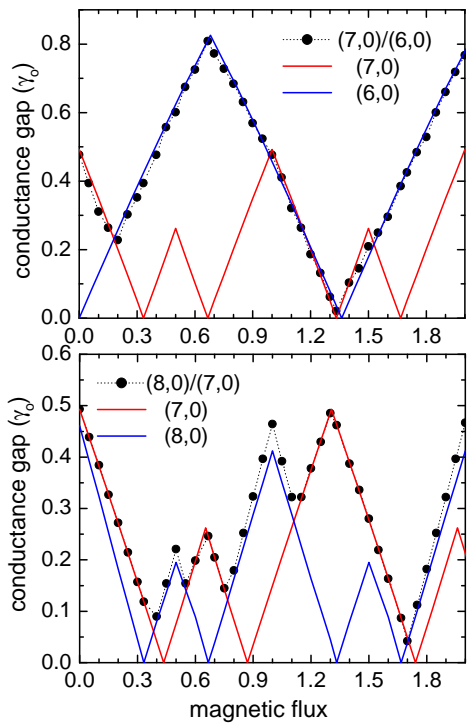}
\end{center}
\caption{''Color online''. Conductance gap size as functions of the
magnetic flux for the S/M [(7,0)/(6,0)] and the S/S [(8,0)/(7,0)]
junctions. The conductance gap for the pristine component tubes,
composing each one of the CNH's, are also plotted.}
\end{figure}
The behavior of the electronic properties of the CNH's are analyzed
when the magnetic field is turned on, increasing up to one quantum
flux. It is clear that, once  the magnetic field is turned on, the
energy gap is strongly affected, reflecting the sensitivity of the
junctions to the presence of field. For magnetic fluxes equal to 1/3
and 2/3 $\phi$/$\phi_o$, the LDOS of both studied junctions shows a
metal-insulator-like electronic transition. However, as one can see
in figure 2(b), those states close to the Fermi energy do not
contribute to the electronic conductance of the system. This is
because they are quasi-localized states and not resonant states
(which was verified by means of a detailed sweeping of the LDOS in
the vicinity of the junction). The presence of the topological
defect produces a strong mismatch between right and left electronic
wave functions, giving a null transmission probability through the
junction, even in the presence of the magnetic field.

\begin{figure}[h!]
\begin{center}
\includegraphics[width=8.5cm]{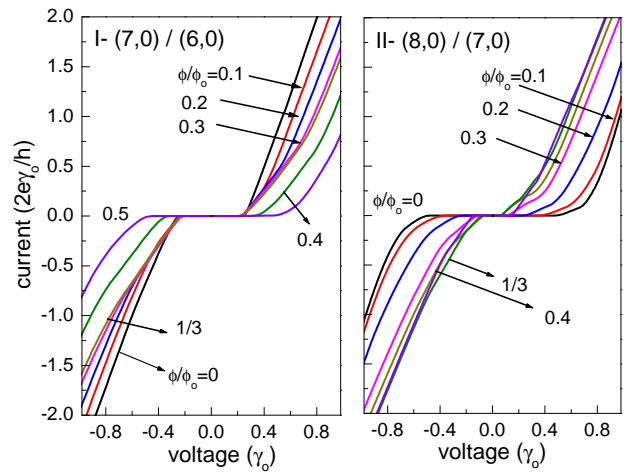}
\end{center}
\caption{''Color online''. Current versus bias voltage for two types
of single junctions: I.- Semiconductor/Metal junction [(7,0)/(6,0)]
and II.- Semiconductor/semiconductor junction [(8,0)/(7,0)] for
different values of magnetic flux, from 0 to 0.5 quantum flux.}
\end{figure}

For helping to understand the behavior of the CNH electronic
conductance on the magnetic field, we explicitly plot in Fig.3 the
conductance gap as a function of a magnetic flux, for the different
studied junctions. In all cases the magnetic flux considered has
been calculated using the area of the biggest tube composing the
CNH's. Actually, the smaller tubes feel a renormalized flux, scaled
by a factor $\mu =(R_n/R_{n-1})^2$, with $R_n$ being the (n,0)tube
radius. Localized states within the gap have been ignored as they
may be viewed as states of null width. A well-known result concerned
to pristine straight tubes under magnetic field is that they
exhibit, metal-insulator-like transitions. Metallic tubes open gaps
as soon as a magnetic flux starts threading it (the maximum gap
occurring at a half of a quantum flux). On the other hand, a
semiconducting tube closes its intrinsic gap at 1/3 and 2/3 of a
quantum flux\cite{Ando2}. An Aharonov Bohm (AB)-type effect for the
gap size has been predicted and may be written in terms of the tube
gap at null field\cite{Lu}. The possibility of similar AB effect
occurrence for the studied CNH's was not been observed. For both
CNH's (S/S and S/M) it is found that the conductance gap of the
structures is given essentially by the biggest gap of the two
constituent tubes, which evolves with the magnetic flux. Similar
results have been obtained for other S/S and M/S junctions.

The characteristic curves of current versus applied bias for the
studied S/M and S/S junctions are shown in Fig.4, for different
magnetic flux intensities. The current across the junction is
obtained via the integration of the transmission function, taking
into account the Fermi distribution of both leads and assuming that
the total potential drop along the heterostructure (bias equal to V)
was fully restricted to the junction extension, linking both
tubes\cite{Faraj}. Within this scheme, an extra potential energy was
added in the diagonal term (site energy) of the tight binding
Hamiltonian.  Alternatively, one may use the Keldysh formalism based
on non-equilibrium Green functions\cite{Keldysh,Carioli,Meir} to
calculate the transport properties in the non-linear regime.
Actually, both treatments are equivalent for coherent transport. For
zero magnetic flux the current gap size of the CNH's is given by the
mean value of the gap energies of the individual tubes. Ohmic curves
are then not expected for this type of nanotube heterostructure
which always involves a semiconducting part. It is easy to show that
the minimum gap size is achieved for a M/S structure, being equal to
$V_{gap}=\epsilon_g/2$, with $\epsilon_g$ the semiconducting gap
energy. Besides the band energy shift imposed by the applied
voltage, a further confinement effect is noticed when a magnetic
flux threads the structure.

\begin{figure}[h!]
\begin{center}
\includegraphics[width=7.5cm]{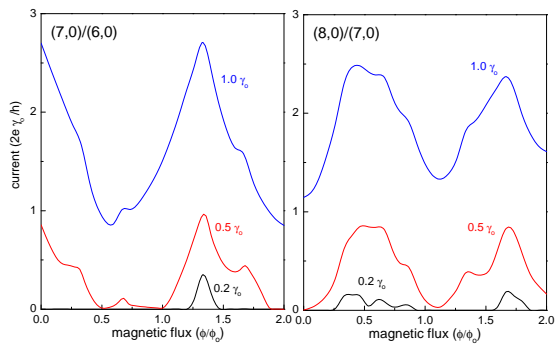}
\end{center}
\caption{''Color online''. Current as a function of the magnetic
flux for different applied bias. Right and left panels correspond to
the (7,0)/(6,0) and (8.0)/(7,0) CNH's, respectively. }
\end{figure}
As it is well known, the field lifts the degeneracy of the
electronic states. However, differently from the case of pristine
tubes, where the semi-classic electronic orbits are split leading to
periodic constructive and destructive quantum interference
phenomena, the CNH's do not exhibit oscillatory current behavior.
The dependence of the current on the magnetic flux for fixed
voltages may prove this issue, as it is shown in Fig.5. The AB
period given by one quantum flux, observed in pristine CN's, is
destroyed. Otherwise, general features marked by peaks and valleys
in the current curves are preserved for different bias intensities,
although exhibiting higher current values as the bias voltage
increases. Also remarkable are the differences between the current
behavior of both considered CNHs, mainly concerned to the magnetic
flux range in which they exhibit a metallic or a semiconducting
character. All theses points emphasize the role played by magnetic
and electric fields on modifying the physical properties of nanotube
structures which may be used to manipulate properly their responses
and potential uses in nanoelectronics.

\subsection{Nanotube quantum dot}

Heterojunctions, such as the discussed ones (n,0)/(n-1,0) are now
put together forming  a nanotube based quantum dot. The size of the
dot is defined by the number of rings forming the internal tube
given here by the integer N, $(n,0)/(n \pm 1,0)_{N}/(n,0)$. Results
for the conductance of $(6,0)/(5,0)_{N}/(6,0)$ CNQD in the absence
of a magnetic flux, are presented in Fig.6, considering dots
composed of 2, 6 and 10 rings in the internal part, within the
leads.
\begin{figure}[h!]
\begin{center}
\includegraphics[width=6.5cm,height=6.5cm]{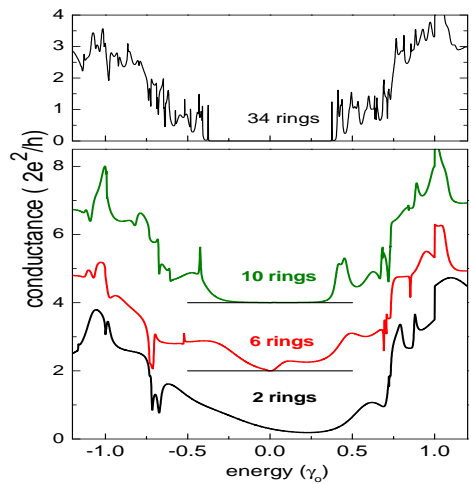}
\end{center}
\caption{''Color online''. Conductance as function of the Fermi
energy of (6,0)/(5,0)$_N$/(6,0) CNQD's of different length: N=2,
(black on-line), N=6 (red on-line), and  N=10 (green-olive
on-line)). The conductance curves for N=6 and N=10 cases were
displaced 2 units of quantum conductance for a better visualization.
In the upper panel N=34 rings.}
\end{figure}
In the upper panel, we show the conductance for a CNQD with N=34
rings. Similarly to the case of single junctions, the electron-hole
symmetry is lost for the quantum dot structure due to the
topological defects at the interfaces of the (5,0) and (6,0) tubes.
As expected, a clean gap is obtained, similar to the semiconducting
gap of the (5,0) pristine CN. The CNQD conductance-gap decreases
with the size of the dot. In this example, the minimum value for the
gap has been obtained for N=6, in which case zero conductance is
only achieved at the Fermi energy. Applying a small gate voltage a
metallic behavior for the CNQD can be obtained. In particular, for a
N=2 dot (length equal to $2a_{cc}$), there is a non-null conductance
at the energy region close to the Fermi level due to the overlap
between the metallic wave functions of the leads across the small
dot region.

\begin{figure}[h!]
\begin{center}
\includegraphics[width=7.0cm]{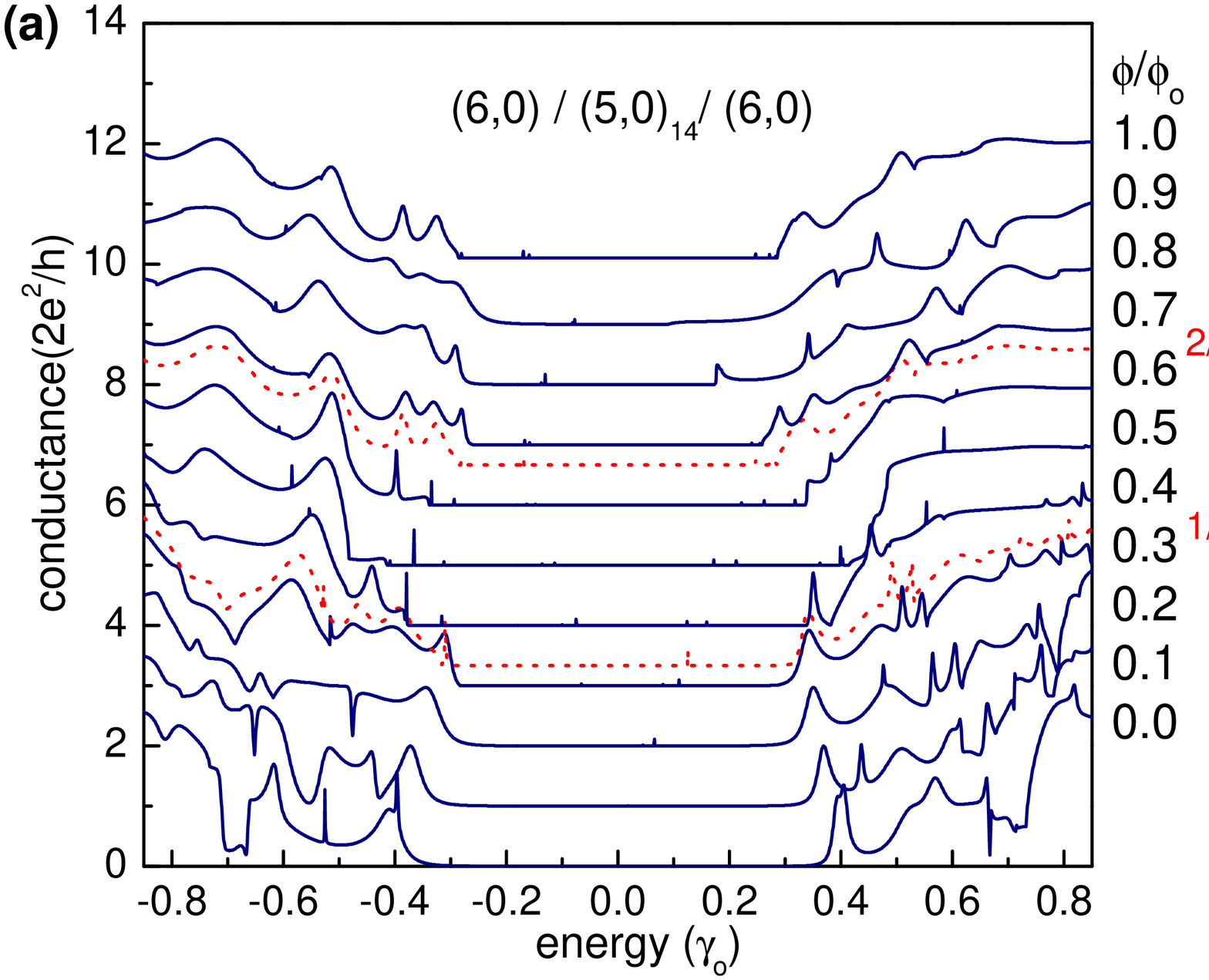}
\includegraphics[width=7.0cm]{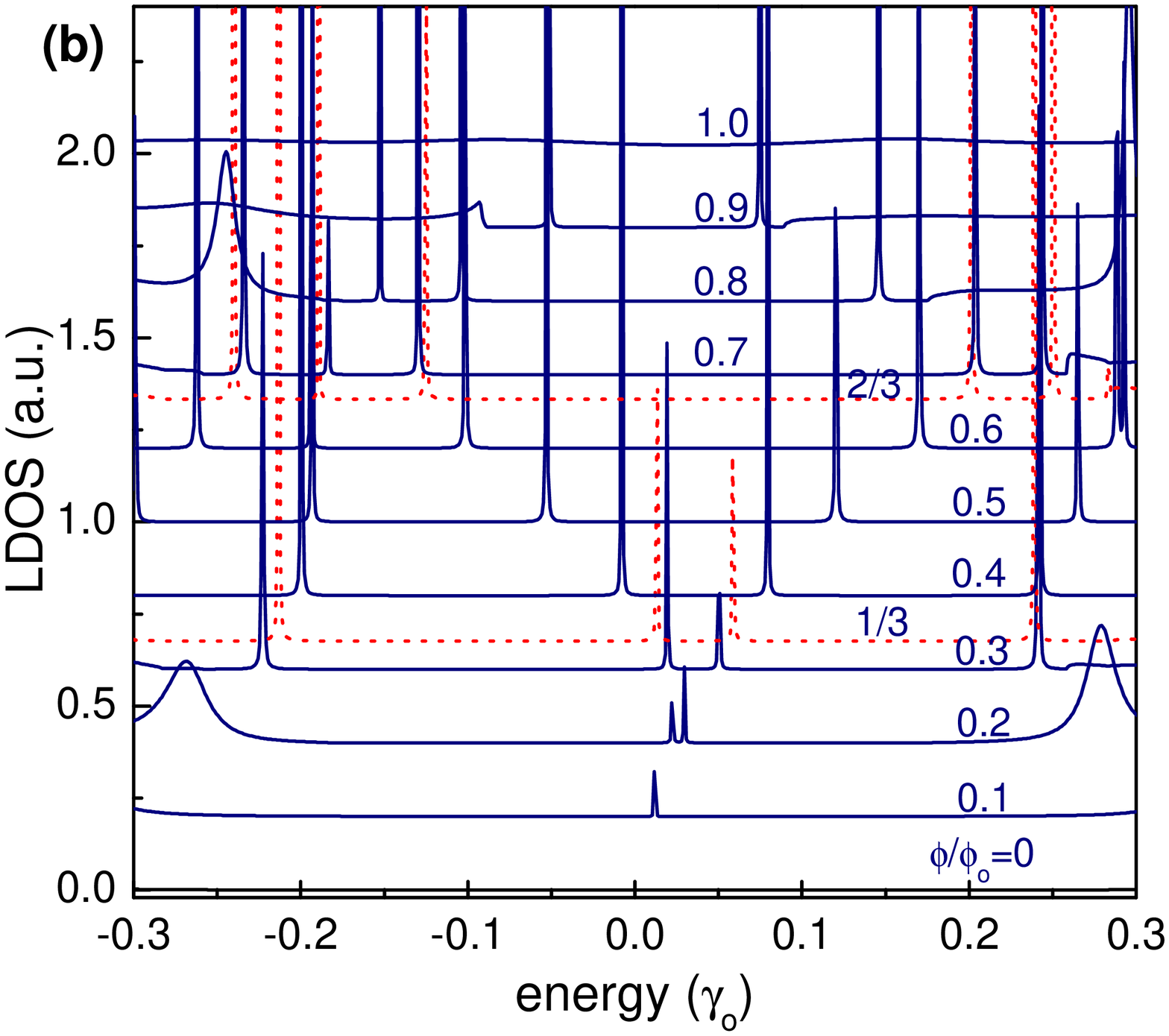}
\end{center}
\caption{''Color online''. (a) Conductance behavior as a function of
the energy  and (b) LDOS in the energy region close to the Fermi
level, for magnetic fluxes up to 1.0 $\phi / \phi_o$ }
\end{figure}

The dependence of the conductance gap on the magnetic field, for a
semiconducting (6,0/(5,0)$_{14}$/(6,0) CNQD is shown in Fig. 7(a).
For this particular dot, a conductance gap is always present for
different  magnetic fluxes, in contrast to the case of a pristine
(5,0) tube for which the gap closes  for magnetic fluxes equal to
1/3 and 2/3 $\phi_o$  (dashed curves in the figure). Usually, two
kind of states can be distinguished for CNQDs: interface and
resonant states. The nature of such states may be determined, for
instance, by performing an accurate sweeping of the LDOS within the
dot region. This region includes the interface defective rings and
also a few rings within the leads\cite{Chico1,Rocha}.
\begin{figure}[h!]
\begin{center}
\includegraphics[width=7.00cm]{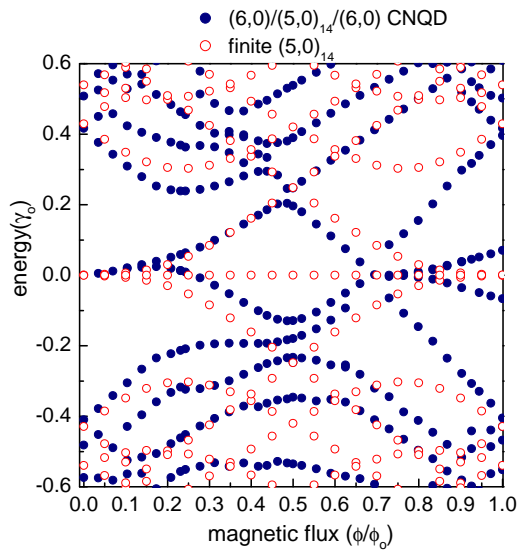}
\end{center}
\caption{''Color online''. Electronic states evolution with the
magnetic flux intensities for a (6,0)/(5,0)$_{14}$/(6,0) CNQD (dark
circles) and for a finite (5,0) CN - without coupled leads, formed
by 14 rings (light symbols). }
\end{figure}
Interface states are typically localized within the energy range
close to the Fermi level and have a weak dependence on magnetic
field intensity. The resonant states are at energies out of the CN
gap region ($< 0.4 \gamma_o$ in the present example, corresponding
to the approximate gap energy at zero magnetic field, for a (5,0)
tube). For null field, this particular CNQD structure does not
exhibit localized levels in the gap region, as can be seen in the
LDOS displayed in figure 7(b). However, as the field is turned on,
the LDOS clearly shows that localized states appear  close to the
Fermi energy, and that they oscillates as a function of the magnetic
field.

The explicit dependence of the energy states on the magnetic flux is
shown in Fig. 8, with full dot symbols. We have also added, for
comparison, the results for a finite short (5,0) CN, composed of 14
rings (empty dots). One should noticed that in this case the results
are presented as a function of the magnetic flux threading a (5,0)
tube, to allow the comparison, whereas for the LDOS exhibited in
fig. 7(b) the used magnetic fluxes correspond to the bigger (6,0)
tube. The finite tube energy spectrum shows states with energies
near the Fermi energy corresponding to edge states whose
wavefunctions are localized at both ends of the finite
tube~\cite{Sasaki}. As expected, the edge state pinned at the Fermi
energy, appearing for the finite CN, is not present in the energy
spectrum of the infinite quantum dot heterostructure. The
magnetic-field induced defect states of the CNQD, shown in figure 7
(b) as the peaked structures in the LDOS,  present quite similar
dependence on the magnetic field as the states of the finite tube.
They may be considered as edge-like states that are lift by the
magnetic field due to the extra confinement imposed by the field. It
is also evident the lack of electron-hole symmetry in the spectra of
the CNQD  as the magnetic field increases, due to the presence of
topological defects. Depending on the length and diameter of the
dot, and also on the atomic details of the junction, different
constituent tubes may be matched, forming different CNQDs. These
geometrical aspects will dictate the presence and nature of the
electronic states which, in turn, may be modulate by the magnetic
flux threading  the structures.

\section{Summary}
We have calculated local density of states and conductance of
different heterostructures (single junctions and nanotube quantum
dots) under the influence of magnetic fields. Emphasis was put on
analyzing the gap modulation induced by the magnetic flux threading
the structures and how the geometric details of the individual tubes
composing the systems may affect the transport responses.
Differently from the pristine tubes, the conductance gaps of the
studied CNHs do not exhibit AB-like periodic oscillations as a
function of the magnetic flux.  This lack of periodicity was also
found for the characteristic curves current versus voltage. By
comparing the field dependence of the states of the dot structure
with the corresponding states of the finite tube (central part of
the dot) we were able to identify the nature of particular
electronic states appearing in the energy range close to the Fermi
level. The used theoretical approach incorporates the atomic details
of the topological defects by performing an energy relaxation via
Monte Carlo calculation. A theoretical treatment taking into account
the charge fluctuations imposed by external bias is presently being
studied. We believe that this kind of theoretical studies on
nanotube composed systems, together with the fast development on new
synthesis techniques, should help to control and modulate the
physical responses of these nanostructures.

\section{Acknowledgments} This work was supported by the Brazilian Agencies CNPq,
CAPES, FAPERJ, Instituto do Mil\^ enio, and PRONEX-CNPq-FAPERJ grant
171.168-200, also by the Iniciativa Cientifica Milenio P02-054-F,
Fondecyt 1050521 and 7060290, and Andes Foundation under the project
C-14055/20, and by the Science Foundation Ireland (SFI).

\end{document}